\documentclass[
 aip,
 rsi,
 reprint,
 floatfix,
 superscriptaddress,
 amsmath,amssymb,
]{revtex4-2}
\usepackage{graphicx}
\usepackage{dcolumn}
\usepackage{bm}
\usepackage[utf8]{inputenc}
\usepackage[T1]{fontenc}
\usepackage{mathptmx}
\usepackage{xcolor}
\usepackage{booktabs}
\usepackage{overpic}

\begin{document}
\title{Actively stabilized Fabry--P\'erot notch filters for correlated photons from cold two-level atoms}

\author{G. C. Borges}
\email{gabriel.cborges@ufpe.br}
\affiliation{Departamento de F\'isica, Universidade Federal de Pernambuco, 50670-901 Recife, Pernambuco, Brazil}

\author{M. Duclerc}
\affiliation{Departamento de F\'isica, Universidade Federal de Pernambuco, 50670-901 Recife, Pernambuco, Brazil}

\author{G. Palacios}
\affiliation{Departamento de F\'isica, Universidade Federal de Pernambuco, 50670-901 Recife, Pernambuco, Brazil}
\affiliation{Centro de Ci\^encias Exatas e Sociais Aplicadas, Universidade Estadual da Para\'iba, 58706-550, Patos, PB, Brazil}

\author{A. A. C. de Almeida}
\affiliation{Departamento de F\'isica, Universidade Federal de Pernambuco, 50670-901 Recife, Pernambuco, Brazil}

\author{N. R. de Melo}
\affiliation{Departamento de F\'isica, Universidade Federal Rural de Pernambuco, 52171-900, Recife, Pernambuco, Brazil}

\author{D. Felinto}
\affiliation{Departamento de F\'isica, Universidade Federal de Pernambuco, 50670-901 Recife, Pernambuco, Brazil}

\begin{abstract}
We report two actively stabilized Fabry--P\'erot cavities operated as reflection-mode notch filters for biphotons generated by spontaneous four-wave mixing in a cold two-level atomic ensemble. In this source, elastic scattering at the excitation frequency increases the single-detector count rates and the accidental-coincidence background, thereby limiting the observed nonclassical correlations. One cavity is placed in each photon-collection mode. An acousto-optic modulator frequency-dithers an auxiliary field, and synchronous demodulation of the transmitted power provides the locking discriminator. Light resonant with the auxiliary field is transmitted, whereas the detuned components are reflected toward the photon detectors. By tuning the auxiliary field to the same frequency as the excitation field, we filter these components and direct a larger portion of the quantum-correlated sidebands to the detection. The cavities have linewidths of approximately $2~\mathrm{MHz}$ and remain locked for longer than a typical correlation acquisition. The measured Cauchy--Schwarz parameter exceeds unity in both configurations, with a higher central value with filtering ($\overline{R} = 4.15 \pm 0.93$) than without ($\overline{R} = 2.33 \pm 0.20$). This architecture provides actively stabilized rejection of a narrow spectral interval, simultaneously, for two fields detected in different directions. Filter performance could be further improved by adjusting the cavity linewidth to better match the spectral width of the central component and by reducing losses in the filters.
\end{abstract}

\maketitle

\section{Introduction}

Spontaneous four-wave mixing (SFWM) in atomic ensembles is a useful source of narrowband photon pairs for light--matter interfaces and quantum networks \cite{Duan2001,Chou2007,Kimble2008,Azuma2023,Craddock2024}. Such applications typically employ multi-level systems with various degrees of freedom (wavelength, polarization, etc)  to filter quantum correlations from any background noise, allowing for the observation of very high correlations~\cite{Laurat2006} as required by quantum repeater protocols~\cite{Duan2001,Mendes2011}. In 2022, our group reported the observation of quantum correlations in SFWM from cold ensembles of pure two-level atoms in free space, with the quantum correlations prevailing over all background noise without filtering~\cite{PRL2022,Marinho2025}. This indicates the presence of robust and strong quantum correlations in the system, even as this first work reported only small levels of quantum correlations. 

Ensembles of cold two-level atoms in free space can have long coherence times related to their atomic external degrees of freedom~\cite{deAlmeida2016,Capella2022}. These memories based on external degrees of freedom are important components to develop quantum memories using higher order nonlinearities, which could be used to explore quite different photon-number correlations than in SFWM~\cite{deOliveira2015}. Previous work by our group indicates that a six-wave mixing memory would already need to rely on storage of information in external degrees of freedom and readout over a pure two-level transition~\cite{Borba2017}. The level of quantum correlations reported in Refs.~\onlinecite{PRL2022,Marinho2025}, however, is not enough to explore such applications of higher complexity.

\begin{figure}[ht]
\centering
\includegraphics[width=8cm,angle=0]{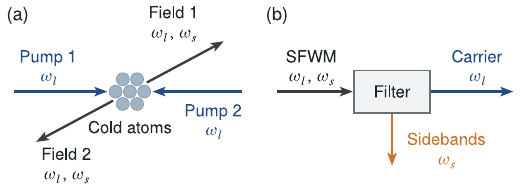}
\vspace{-0.3cm}
\caption{(a) Schematic view of a spontaneous four-wave mixing process. During the measurement window, two counterpropagating pump beams at frequency $\omega_l$ excite the atomic cloud. The scattered light is collected in two conjugate spatial modes, labeled 1 and 2. The scattered light has components with the carrier frequency $\omega_l$ and sidebands of frequency $\omega_s$. (b) Notch filter for the scattered light, separating sidebands from the carrier frequency.}
\label{fig:intro}
\end{figure}

In order to move forward developing applications involving pure two-level systems, it is necessary then to introduce some filtering to enhance their quantum correlations. For these systems, the quantum correlations are concentrated in the sidebands of the Mollow triplet~\cite{Mollow1969}, and their extraction through filtering has been suggested~\cite{CohenTannoudji1979} and first tried~\cite{Aspect1980} more than forty years ago. However, the observation and distillation of these quantum correlations proved to be quite challenging. On the one hand, its small unfiltered value was hard to observe by itself, as common imperfections in the system would typically decrease the quantum correlations down to classical levels~\cite{Marinho2025}. On the other hand, since frequency is the only degree of freedom we have for filtering in pure two-level systems, the proximity of the carrier frequency $\omega_l$ to its sideband frequencies $\omega_s$ imposes stringent requirements on any frequency filter, see Fig.~\ref{fig:intro}(a) for our typical geometry of SFWM. 

For the Mollow triplet, these sidebands occur at $\omega_s = \omega_l \pm \Omega_g$, with $\Omega_g$ the generalized Rabi frequency of the system. For large detunings $\Delta$ of $\omega_l$ with respect to the atomic resonance, we have typically $\Omega_g \approx \Delta$, a condition that will hold for all data presented here and for the previous publications of our group on the problem~\cite{PRL2022,Marinho2025}. To separate the sidebands from the carrier frequency, we should then employ a band-pass (or notch) filter that directs $\omega_l$ and $\omega_s$ to different outputs, as schematically shown in Fig.~\ref{fig:intro}(b).  The bandwidth of the filter should be much smaller than the applied detuning.

The problem of distilling the quantum correlations in a Mollow-triplet spectrum is common to various systems exploring two-level emitters, such as in linear ensembles of cold atoms coupled to a nanofiber~\cite{Prasad2020,Cordier2023}, or a single atom in free space~\cite{Masters2023}, or a semiconductor quantum dot coupled to a photonic waveguide~\cite{Liu2024}. References~\onlinecite{Prasad2020,Cordier2023} excited the atoms in resonance and used the combination of strong directionality and absorption in the sample to eliminate the excitation frequency, transmitting mainly the out-of-resonance sidebands. On the other hand, references~\onlinecite{Masters2023,Liu2024} apply single notch filters to separate the sidebands from the excitation frequency right at the output. The filter employed in Ref.~\onlinecite{Masters2023} is based on a passively-stabilized fiber-ring resonator~\cite{Maron2026}, while Ref.~\onlinecite{Liu2024} used a combination of an actively stabilized Fabry-Perot filter with a volume phase holographic transmission grating~\cite{Liu2024B}. These were all quite different solutions adapted for the peculiarities of the Mollow spectra in each system.

For our system, the first important difference is the need for pairs of filters, as our quantum correlations appear in the cross signal of a SFWM setup, see Fig.~\ref{fig:intro}(a). We also do not have the directionality of the 1-dimensional system of Refs.~\onlinecite{Prasad2020,Cordier2023} and cannot use absorption to filter the excitation light while still keeping the sidebands. Our first approach was to use two passively-stabilized Fiber Fabry-Perot filters, positioned symmetrically around $\omega_l$ at the centers of the sidebands~\cite{OL2023}. In this way, we were able to enhance the quantum correlations by roughly a factor of two with respect to our best unfiltered results previously reported, as measured from the amount of violation of a Cauchy--Schwarz inequality. These were broadband filters, of around 600~MHz bandwidth, with good passive stability but requiring large values for $\Delta$. The need to work far away from resonance ended up limiting further improvements of this filter, as the signal-to-noise ratio decreased with the signal itself decreasing for such large detunings. 

Here we report our second generation of filters, consisting now of actively-stabilized narrowband notch filters tuned to transmit the excitation frequency in both of the conjugated signals, while reflecting the sidebands. They employ a single free-space Fabry-Perot filter for each detected field and can operate both near and far from resonance, limited only by the Free Spectral Range (FSR) of the filters, on the order of tens of GHz. With these filters, we already achieved similar performance as reported in Ref.~\onlinecite{OL2023}, but now without the limitations of having to work only with very large detunings.

In the following, we introduce and characterize our Fabry-Perot filters in Sec.~\ref{FPfilters}, including a description of their electronic control of the optical and mechanical degrees of freedom. Section~\ref{sec:sfwm} includes a description of our apparatus for generation and characterization of photon pairs from SFWM. In Sec.~\ref{sec:sfwm}, we also report our first results employing the new filters on the SFWM signal. Section~\ref{conclusion} provides our conclusions and perspectives to continue improving the setup.

\section{Narrowband Fabry--P\'erot filter and transmission lock}
\label{FPfilters}

A narrow reflection-mode notch offers a complementary strategy to the usual selection of photons through a cavity-transmission peak. When the cavity is resonant with the excitation frequency $\omega_l$, light within the central spectral interval exits through transmission, while the detuned components $\omega_s$ return through reflection. In our configuration, the same transmission port provides the locking discriminator and the reflection port provides the filtered photon output, with one independent feedback loop in each of the two collection modes of Fig.~\ref{fig:intro}(a). The experimental challenge is to keep the narrow notch filter centered on the excitation field without allowing the brighter locking reference to reach the single-photon detectors.

In the following subsections, we describe our actively stabilized Fabry--P\'erot notch filter. The cavity length is locked with an auxiliary beam whose frequency is dithered by a single-pass acousto-optic modulator (AOM). In contrast to standard Pound--Drever--Hall (PDH) locking, the discriminator is obtained by synchronous demodulation of the cavity transmission; the reflection port is reserved for the filtered photons. The two feedback loops share an electronic time reference but control their respective cavities independently, and the auxiliary optical fields are interrupted during photon counting. We characterize the cavity response and long-term operation and then, in the following sections, test the filters directly in the SFWM correlation experiment.

\subsection{Fabry--P\'erot cavity}\label{sec:fpc}
A high-finesse Fabry--P\'erot (FP) cavity, composed of two highly reflective mirrors separated by a distance $L$,
supports discrete longitudinal modes defined by the resonance condition $2kL = 2\pi q$, where $k = 2\pi/\lambda$ and $q$ is an integer. The corresponding resonant frequencies are equally spaced by the free spectral range, $\mathrm{FSR} = c/2L$. For the present purpose, the relevant measured quantities are the FSR, the full width at half maximum (FWHM) of the transmission peak, and the depth of the reflection minimum. The finesse is defined as $\mathcal{F}=\mathrm{FSR}/\Delta\nu_{\mathrm{FWHM}}$. Because the reflected response also depends on intracavity losses and mode matching, we characterize it directly from transmission and reflection scans rather than from the nominal mirror reflectivities.

The cavity geometry and operating parameters are summarized in Table~\ref{tab:filter_parameters}. Within the calibration resolution, the two devices have comparable FSRs and FWHMs, corresponding to a finesse of order $10^4$. The symmetric, near-planar geometry supports a stable fundamental Gaussian mode compatible with the fiber-coupling optics. Its compactness facilitates integration into the detection path, while its sensitivity to alignment and cavity-length fluctuations of the narrow resonance motivates passive isolation and active length control. Because a length fluctuation $\delta L$ shifts the resonance by $\delta\nu\simeq-(\nu/L)\delta L$, active stabilization is required to keep the narrow rejection feature centered on the Rayleigh component.

\begin{table}[ht]
\centering
\caption{Operating parameters relevant to the filtering channels.}
\label{tab:filter_parameters}
\begin{tabular}{@{}lll@{}}
\toprule
Quantity & Value & Scope \\
\midrule
Cavity length & $\sim7~\mathrm{mm}$ & both cavities \\
Mirror curvature & $250~\mathrm{mm}$ & both mirrors in each cavity\\
Free Spectral Range & $\sim21.4~\mathrm{GHz}$ & both cavities \\
Measured FWHM & $\sim2~\mathrm{MHz}$ & both cavities \\
Narrow-field rejection & $\sim70\%$ & both cavities \\
Frequency-dither rate & $10~\mathrm{kHz}$ & shared electronic reference \\
Reference interruption & $2~\mathrm{ms}$ & per experimental cycle \\
\bottomrule
\end{tabular}
\end{table}

Each Fabry--P\'erot cavity consists of two highly reflective curved mirrors mounted on a rigid opto-mechanical structure. The setup is assembled on a mechanically stable baseplate and enclosed within an acrylic housing, which provides passive isolation against acoustic noise and air flow. These measures help reduce mechanical vibrations and thermal drifts. One of the cavity mirrors is mounted on a precision translation stage, allowing fine adjustment of the cavity length and alignment during the initial setup. The second mirror is attached to a piezoelectric transducer (PZT), which enables dynamic control of the cavity length. The same actuator is used for calibrated resonance scans and for corrections from the feedback loop. The cavity response is characterized by scanning its length while recording transmission and reflection. The frequency scale for each device is calibrated using known hyperfine separations from saturated-absorption spectroscopy, yielding the comparable FSRs and linewidths reported in Table~\ref{tab:filter_parameters}.

\subsection{AOM-based transmission lock}\label{sec:pdh}

\begin{figure}[hbt]
\centering
\includegraphics[width=\columnwidth]{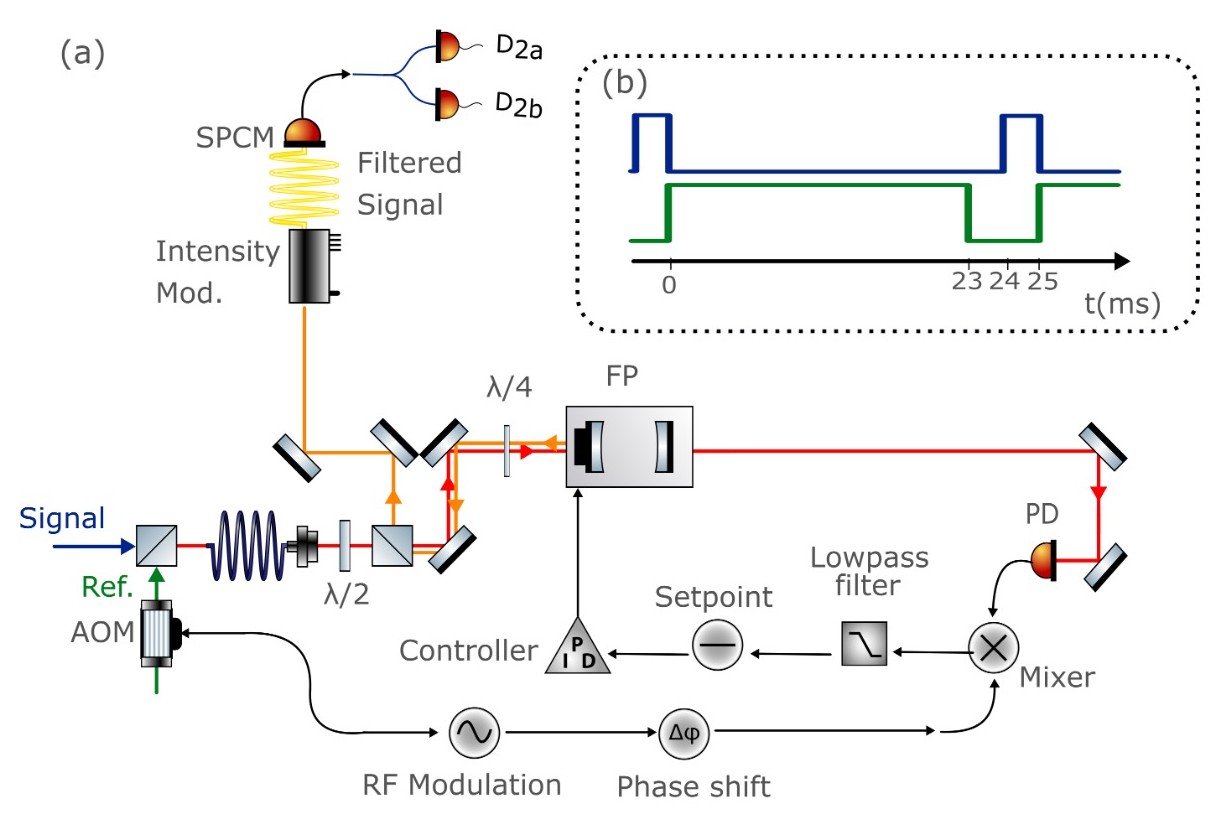}
\vspace{-0.8cm}
\caption{(a) Reflection-mode Fabry--P\'erot filter and transmission-locking loop. The SFWM field and auxiliary reference share a single-mode fiber and are projected onto the same cavity polarization before the Fabry--P\'erot cavity. The AOM frequency-dithers the reference in a single pass. The transmitted power is synchronously demodulated, and the PID output drives the cavity PZT. Off-resonant photon components collected in reflection pass through an intensity modulator before detection by two single-photon counting modules. The mixer, low-pass filter, phase shift, and controller blocks are a functional schematic; the corresponding signal processing is implemented digitally in an FPGA-based lock box. (b) Timing sequence for alternating reference and photon-counting intervals. PBS, polarizing beam splitter; PD, photodiode; SPCM, single-photon counting module.}
\label{fig:setup_nbf}
\end{figure}

The locking method is inspired by PDH stabilization, in which a modulated field provides an antisymmetric frequency discriminator around a cavity resonance \cite{Drever1983,Black2001}. Dither locks based on a transmission signal are well established \cite{Liu2020,Zugenmaier2018}, and AOM-based modulation has also been used in PDH implementations \cite{Zeng2021}. Our implementation combines these elements: the optical frequency is dithered with an AOM, and the discriminator is obtained from transmitted rather than reflected power. We therefore refer to it as an AOM-based frequency-dither transmission lock, rather than as a standard PDH lock. The control loop is schematically shown in Fig.~\ref{fig:setup_nbf}(a).

A triangular waveform at $10~\mathrm{kHz}$ is applied to the frequency-modulation input of the radio-frequency driver for the $200~\mathrm{MHz}$ AOM. The mean auxiliary frequency incident on the cavity is adjusted to the Rayleigh component. The transmitted photodiode signal is synchronously demodulated at the dither frequency by an FPGA-based lock box, producing a discriminator with a zero crossing at the transmission maximum. The demodulated signal is processed digitally and fed to a proportional--integral--derivative (PID) controller that drives a piezoelectric transducer (PZT) on one cavity mirror. The cavity is thus locked to the reference laser, while its reflected port remains available for the quantum field.

\subsection{Optical paths and polarization control}
\label{subsec:Setup_filter}

The filter receives the weak SFWM field and an auxiliary reference used only for stabilization. Figure~\ref{fig:setup_nbf}(a) shows their common optical path and the separation of the transmitted locking signal from the reflected photon channel. Both fields are coupled into the same single-mode fiber before entering the filter, providing spatial-mode overlap at the cavity input and facilitating coupling to the fundamental cavity mode. The auxiliary reference supplies the optical power required for a reliable discriminator; the SFWM field alone is too weak for this purpose. Because the brighter reference would contaminate the photon counts, a mechanical chopper blocks it during the measurement window.

The reference and photon fields therefore share the spatial mode of the filter but are separated in time at the detectors. This arrangement removes the need to derive the locking signal from the generated photons. Reflection-mode operation also avoids tuning a narrow transmission resonance to either inelastic component, making the locking performance largely independent of the photon-production rate. As represented in Fig.~\ref{fig:setup_nbf}(b), the reference is absent for $2~\mathrm{ms}$ of each $25~\mathrm{ms}$ experimental cycle, including the $1~\mathrm{ms}$ photon-counting interval. The PID remains enabled throughout the cycle. The sustained time-multiplexed operation reported below shows that the cavity can remain locked under this repeated interruption.

After propagation through the fiber, a half-wave plate and a polarizing beam splitter project the SFWM field and the auxiliary reference onto the same linear polarization before the cavity. The wave plate is adjusted to direct toward the cavity most of the SFWM field and only the fraction of the much brighter reference required for locking. The remaining reference power is discarded at the other PBS port. Both fields therefore address the same polarization mode of the cavity.

The cavity resonance is centered on the auxiliary reference and hence on the pump-frequency Rayleigh component. These resonant fields are transmitted. The frequency-shifted components of the SFWM field are off resonance and are directed from the reflected port to an intensity modulator and a pair of single-photon counting modules, each containing a single-photon avalanche diode (SPAD). Thus, transmission is used for locking and for removing the narrow central component, whereas reflection provides the filtered photon output. The transmitted reference is monitored by a photodiode operating in the linear regime. Its digitally demodulated signal furnishes the frequency discriminator, and the PID output drives the cavity PZT. The reflected photon field is temporally gated by an intensity modulator before being divided between the two detectors assigned to that collection mode. The same sequence is implemented independently in the second arm.

This configuration separates the transmitted locking signal from the reflected filtered field. The cavity is aligned to maximize coupling to the fundamental transverse mode ($\mathrm{TEM}_{00}$), which is essential for efficient filtering and a reproducible transmission discriminator. Once established, the alignment remains stable over the duration of a typical correlation acquisition.

\subsection{Two-cavity locking scheme and electronic control}
\label{subsec:PDH-like_scheme}

The two cavities, FP$_1$ and FP$_2$, are controlled by separate digital lock boxes. The first uses its internal $10~\mathrm{kHz}$ reference for dither and demodulation. The same electronic reference is sent to the second unit, which tracks it with a phase-locked loop before demodulating its own transmitted signal. This arrangement preserves the relative timing of the two discriminators while allowing each PID to control its own PZT. Each cavity has its own transmitted signal, demodulator, PID, and PZT, so the two feedback processes are independent. Sharing the $10~\mathrm{kHz}$ time reference does not couple their control actions, although both channels depend on the common optical reference and can therefore respond to the same laser or laboratory disturbances.

\begin{figure}[ht]
\centering
\includegraphics[width=0.8\columnwidth]{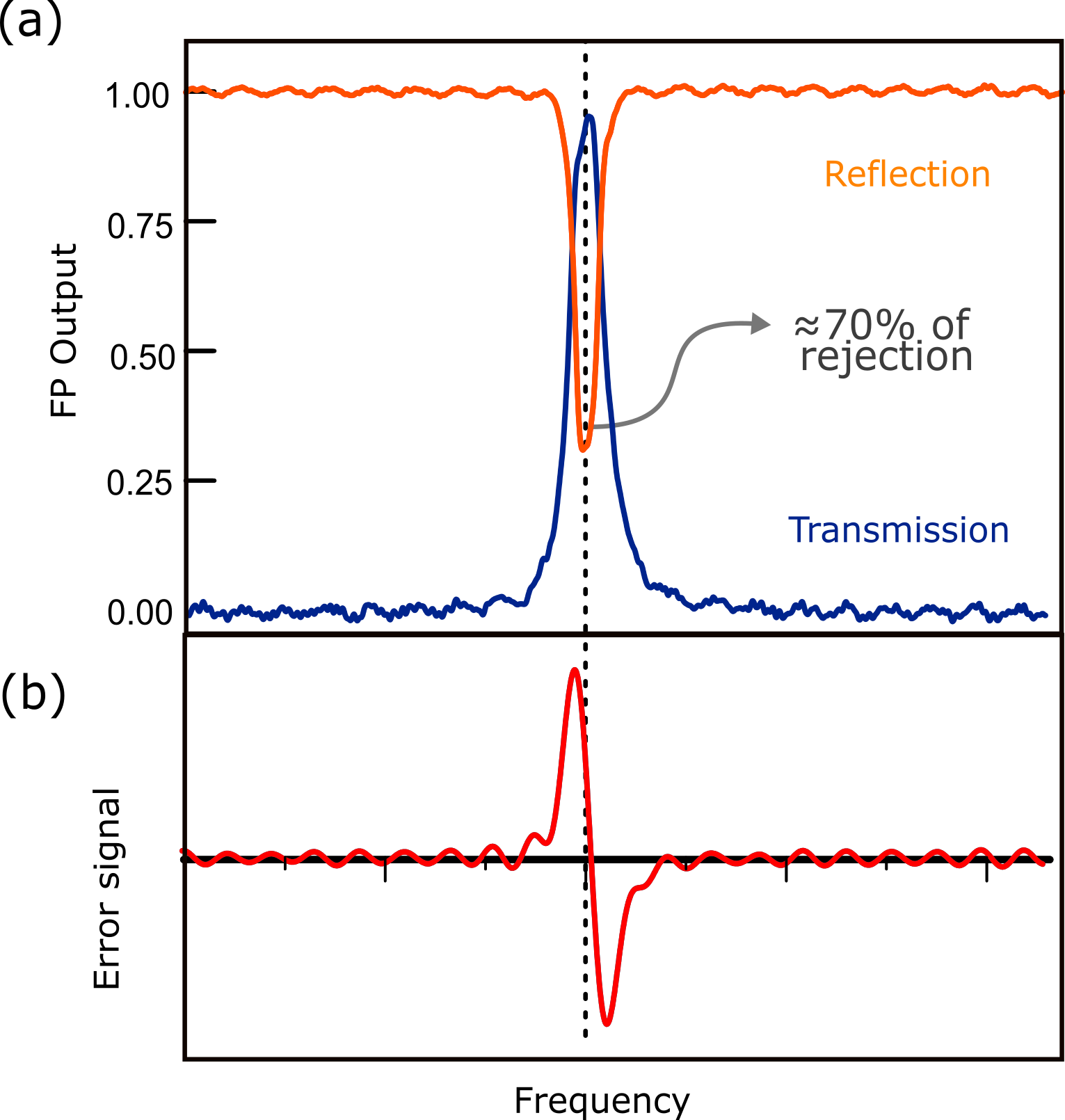}
\caption{Illustrative cavity-locking signals; the curves are not the calibrated data used to determine the FSR or FWHM. (a) Representative reflection (orange) and transmission (blue) traces across a resonance. The approximately $70\%$ reflection dip represents the notch rejection measured with a spectrally narrow field. (b) Representative discriminator obtained by synchronous demodulation of the transmitted photodiode signal; its zero crossing defines the locking point.}
\label{fig:output_fp_pdh}
\end{figure}

For each cavity, a slow scan first locates a transmission resonance, after which the demodulation phase and set point are adjusted around the corresponding zero crossing and the PID loop is closed. Once locked, the controller remains enabled throughout the experimental cycle. During the short interval in which the reference is blocked, however, no optical discriminator is available; normal closed-loop correction resumes when the reference returns. Figure~\ref{fig:output_fp_pdh} illustrates the signals used to operate the lock. The transmission maximum and reflection minimum identify the resonance, while synchronous demodulation of the transmitted signal produces the zero crossing used for feedback. The reflection minimum corresponds to approximately $70\%$ rejection for the spectrally narrow reference field.

\subsection{Control-signal monitoring and operational endurance}\label{stability_analysis}

\begin{figure*}[ht]
\centering
\includegraphics[width=17cm]{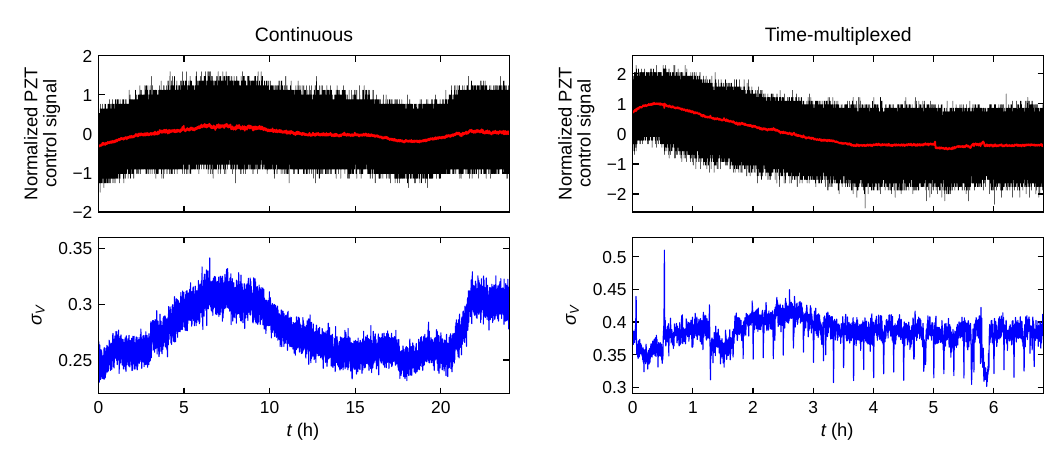}
\vspace{-0.5cm}
\caption{Long-duration records of the normalized PZT control signal. Top: normalized PZT control signal $V(t)$ (black) and moving mean $\mu_V(t)$ (red). Bottom: moving standard deviation $\sigma_V(t)$. Left column: continuous operation for more than $24~\mathrm{h}$, terminated deliberately while the cavity remained locked. Right column: time-multiplexed operation until loss of lock after approximately $7~\mathrm{h}$.}
\label{fig:long_runs}
\end{figure*}

To assess long-term operation, we record a normalized version of the control signal sent to the cavity PZT, denoted by $V(t)$ and displayed in arbitrary units. The traces therefore characterize slow control activity during locked operation rather than the dynamics within an experimental cycle. For visual guidance, the traces include a moving mean $\mu_V$ and a moving standard deviation $\sigma_V$. The first follows the slowly varying operating point, whereas the second highlights changes in the amplitude of the controller corrections. Both quantities were computed using a sliding window with a nominal duration of $20~\mathrm{s}$. This duration is approximately six times as long as the period ($3.2~\mathrm{s}$) corresponding to the lowest selectable characteristic frequency of the feedback controller ($0.3125~\mathrm{Hz}$), and each window contains approximately $10^3$ samples at the acquisition rate of $f_s \approx 50~\mathrm{Hz}$. This choice provides a compromise between statistical robustness and temporal resolution, allowing slow variations in the control signal to be tracked. The left panels in Fig.~\ref{fig:long_runs} show a representative continuous run acquired with the auxiliary reference present and the SFWM source absent. The slow evolution of $V(t)$ indicates changes in the correction applied by the controller. The cavity remained locked throughout this $24\mathrm{h}$ record, with the run terminated deliberately with the cavity still locked. 

In time-multiplexed operation, the chopper blocks the auxiliary field during photon counting while the PID remains enabled. Since the slow monitor does not resolve individual interruptions, the traces with time-multiplexing on the right column of Fig.~\ref{fig:long_runs} characterize samples of the plotted quantities during the time-multiplexed operation rather than the full dynamics within each cycle. The control signals on the top panels do not appear disturbed by the chopped operation. However, $\sigma_V$ exhibits jumps reflecting the perturbation caused by the periods without reference signal feeding the control electronics, which leads to typically shorter locking periods. For the run in Fig.~\ref{fig:long_runs}, the lock was maintained for approximately $7~\mathrm{h}$ before being lost.

The time durations of the locked signals in Fig.~\ref{fig:long_runs} exceed the duration of a typical correlation acquisition. Runs were also interrupted when the common excitation laser was deliberately unlocked for adjustments of the magneto-optical trap. These latter interruptions were not counted as cavity-lock failures.

\section{Spectral filtering of SFWM photons}
\label{sec:sfwm}

\begin{figure}[htb!]
\centering
\includegraphics[width=\columnwidth]{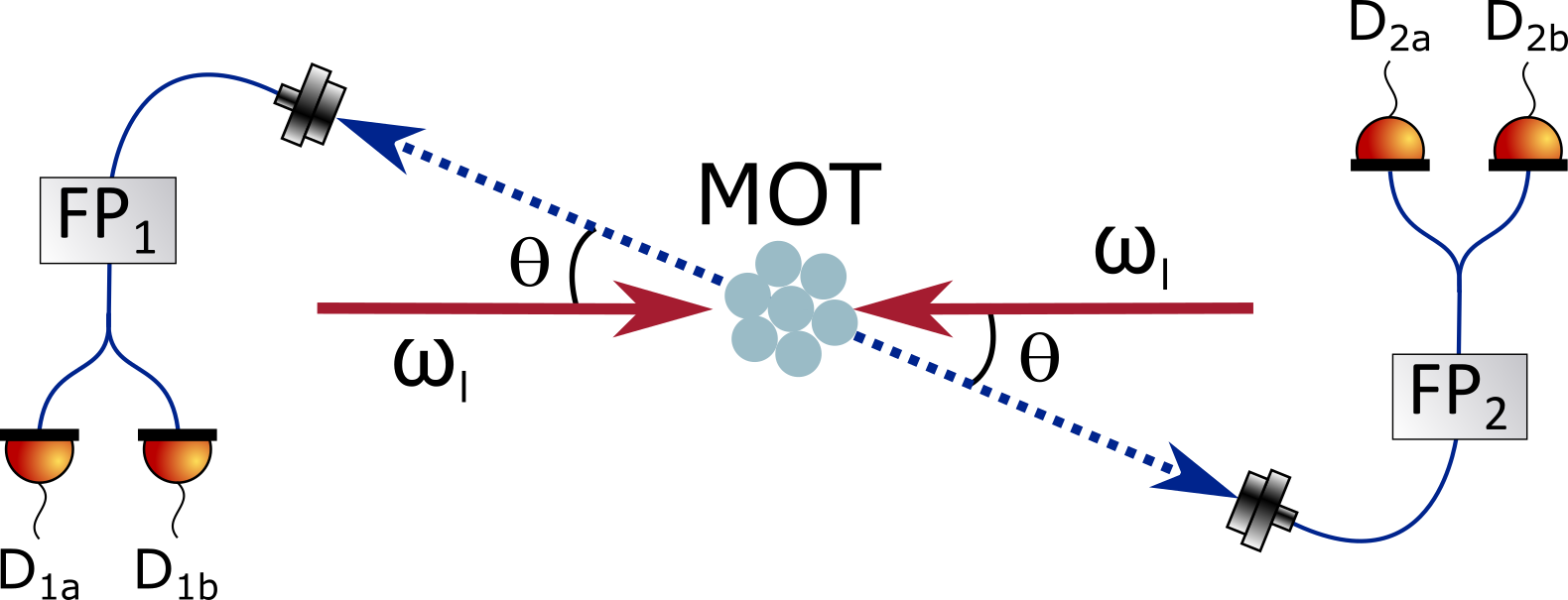}
\caption{SFWM source and detection system. Two counterpropagating pump beams excite the cold atomic ensemble. The scattered light is collected in two spatial modes and sent to independent reflection-mode filters FP$_1$ and FP$_2$, followed by temporal gating and single-photon detection.}
\label{fig:fwm_scheme}
\end{figure}

We validate the filters with the two-level-atom SFWM source used in Refs.~\onlinecite{PRL2022,OL2023}. The source, shown in Fig.~\ref{fig:fwm_scheme}, consists of laser-cooled $^{87}\mathrm{Rb}$ atoms in a magneto-optical trap. Two counterpropagating pump beams drive the detuned transition $\lvert5S_{1/2},F=2\rangle\rightarrow\lvert5P_{3/2},F'=3\rangle$. The circularly polarized beams prepare the atoms predominantly in the stretched ground state $\lvert F=2,m_F=2\rangle$, so that excitation occurs mainly on the closed transition $\lvert F=2,m_F=2\rangle\rightarrow\lvert F'=3,m_{F'}=3\rangle$, providing a pure two-level system.

The experiment is repeated in $25~\mathrm{ms}$ cycles, see Fig.~\ref{fig:setup_nbf}(b). Atomic cooling and trapping occupy most of the cycle, followed by a $1\,\mathrm{ms}$ photon-counting window with the pump beams on. The auxiliary reference is blocked over a slightly longer interval that contains this complete 1\,ms measurement window. Wave plates and polarizing beam splitters select the desired polarization before coupling each collection mode into a single-mode fiber. Each fiber feeds one Fabry--P\'erot filter, whose reflected output is divided between a pair of SPADs. This four-detector arrangement provides two autocorrelation and four cross-correlation functions.

\subsection{Correlation functions}

Arrival times from the four detectors, $D_{1a}$, $D_{1b}$, $D_{2a}$, and $D_{2b}$, are recorded with a time tagger. If $N_t$ is the number of experimental cycles, the single and joint detection probabilities are estimated as
\begin{align}
    p_i(t)&=\frac{N_i(t)}{N_t},\\
    p_{ij}(t,t+\tau)&=\frac{N_{ij}(t,t+\tau)}{N_t},
\end{align}
where $N_i(t)$ counts events in which detector $i$ clicked at the time $t$ of the measurement window. $N_{ij}(t,t+\tau)$ counts events in which detectors $i$ and $j$ clicked at $t$ and $t+\tau$, respectively. The normalized second-order correlation functions are
\begin{equation}
    g^{(2)}_{ij}(t,t+\tau)=
    \frac{p_{ij}(t,t+\tau)}{p_i(t)p_j(t+\tau)}.
    \label{eq:g2}
\end{equation}
To improve the statistics, we average $g^{(2)}_{ij}(t,t+\tau)$ over the whole photon-counting window $\mathcal{W}$ of duration $T=1\,$ms~\cite{PRL2022,OL2023}:
\begin{equation}
    \overline{g}^{(2)}_{ij}(\tau)=
    \frac{1}{T}\int_{\mathcal{W}} g^{(2)}_{ij}(t,t+\tau)\,dt\,.
    \label{eq:g2_average}
\end{equation}
The detector pairs in modes 1 and 2 provide the autocorrelations, while the four pairs connecting the two modes provide the cross-correlations. Constant detection efficiencies cancel from the normalized correlations under otherwise unchanged conditions~\cite{Clauser1974}. However, losses still reduce the number of registered events and therefore degrade the statistical precision, while detector imbalance and background counts are not compensated or subtracted in the data analysis.

For this four-detector configuration, two Cauchy--Schwarz parameters can be formed \cite{Clauser1974}:
\begin{align}
R_1(\tau)&=
\frac{\overline{g}^{(2)}_{2b,1a}(\tau)
      \overline{g}^{(2)}_{2a,1b}(\tau)}
     {\overline{g}^{(2)}_{1b,1a}(0)
      \overline{g}^{(2)}_{2b,2a}(0)},\\
R_2(\tau)&=
\frac{\overline{g}^{(2)}_{2a,1a}(\tau)
      \overline{g}^{(2)}_{2b,1b}(\tau)}
     {\overline{g}^{(2)}_{1b,1a}(0)
      \overline{g}^{(2)}_{2b,2a}(0)}.
\label{eq:R12}
\end{align}
We report their symmetrized value,
\begin{equation}
    \overline{R}(\tau)=\frac{R_1(\tau)+R_2(\tau)}{2}.
    \label{eq:Rbar}
\end{equation}
Classical fields satisfy $R_1\leq1$ and $R_2\leq1$, and hence $\overline{R}\leq1$. Quantum correlations are then demonstrated when $\overline{R}\nleq1$. The filter performance is verified through the comparison of both the correlation parameter and the reduction in the single-detector count rates with and without filtering.

\subsection{Spectral components}

Typically for such two-level systems~\cite{PRL2022,OL2023}, the autocorrelation functions at zero delay will have values of $\overline{g}^{(2)}_{1b,1a}(0), \overline{g}^{(2)}_{2b,2a}(0) \approx 2$, consistent with the expected statistics of a thermal light source~\cite{Loudon1983}. The condition $\overline{R}\nleq1$ implies, then, that functions $\overline{g}^{(2)}_{1,2} (\tau) > 2$ already indicate the presence of quantum correlations in the system. For ensembles of two-level atoms, the cross-correlation function is given by~\cite{Du2007PRL,Wen2007PRA} 
\begin{equation}
\overline{g}^{(2)}_{1,2} (\tau) = 1 + \frac{R_{cc}(\tau)}{R_R^2} \,, \label{g12}
\end{equation}
with $R_{cc}(\tau)$ being the rate of SFWM coincidence counts separated by $\tau$ and $R_R$ the rate of linear Rayleigh scattering in the system, assumed to be the same for both collection channels. Equation~\eqref{g12} implies that quantum correlations will prevail once the rate of coincidences between the two channels is larger than the square of the rate of singles per channel, with $R_R^2$ indicating the rate of accidental coincidences. 

\begin{figure}[htb!]
\centering
\includegraphics[width=8cm]{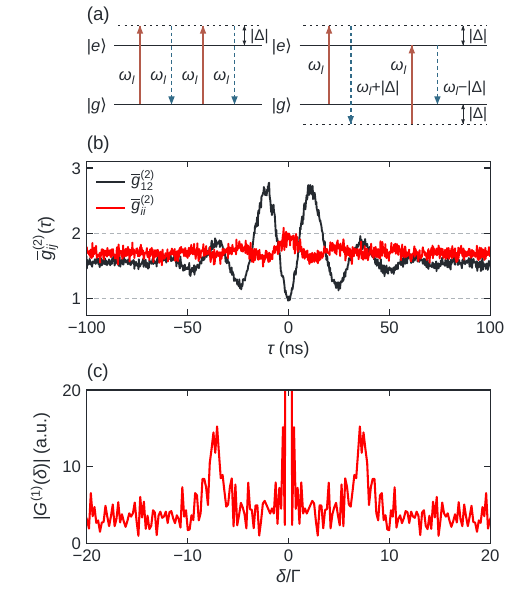}
\vspace{-0.5cm}
\caption{(a) Elastic and inelastic scattering paths in the effective two-level system. (b) Cross-correlation $\overline{g}^{(2)}_{12}(\tau)$ and auto-correlation $\overline{g}^{(2)}_{ii}(\tau)$ (with $i=1,2$) for $\Delta = -6\Gamma$ and without filtering. The horizontal dashed lines mark the uncorrelated level $g^{(2)}=1$ and the reference level $g^{(2)}=2$. For clarity of presentation, the $\overline{g}^{(2)}_{12}$ curve is an average of the four similar curves contributing to $\overline{R}(\tau)$. The $\overline{g}^{(2)}_{ii}(\tau)$ curve is an average of the two similar curves for $i = 1$ and 2. (c) For each collection mode ($i=1,2$), we compute the Fourier transform of $\sqrt{\overline{g}^{(2)}_{ii}(\tau)-1}$ over the delay interval from $-500$ to $+500~\mathrm{ns}$. The red curve shows the average of the magnitudes of the two transforms. The central peak extends beyond the vertical scale shown.}
\label{fig:two_level_spectrum}
\end{figure}

In order to understand the action of the filters extracting the $\omega_l$ component of the scattered light, we need to delve into the spectral components contributing to Eq.~\eqref{g12}. In Fig.~\ref{fig:two_level_spectrum}(a) we show the two processes contributing to the coincidence rate $R_{cc}$. We can have a sequence of two scattered photons with $\omega_s = \omega_l$, or we can have two scattered photons in the sidebands $\omega_s = \omega_l \pm \Delta$. In the first process, the two $\omega_l$ photons can come from different atoms or from the same atom. The contribution from different atoms is the linear Rayleigh scattering contributing to the accidental coincidences of the denominator of Eq.~\eqref{g12}. When the two $\omega_l$ photons come from the same atom, they result in a SFWM process. When the scattered photons have different frequencies ($\omega_l \pm \Delta$), they necessarily come from the same atom in a SFWM process. 

\vspace{-0.0cm}

These three processes result in the experimental $\overline{g}^{(2)}_{1,2} (\tau)$ curve plotted in Fig.~\ref{fig:two_level_spectrum}(b) for $\Delta = -6 \Gamma$, in which $\Gamma / 2\pi \simeq 6.07\,$MHz is the natural linewidth of the excited state. Its plateau in $\overline{g}^{(2)}_{1,2} = 1$ comes from the accidental coincidences, being proportional to $R_R^2$. The oscillating structure on top of the plateau, with minimum at $\tau = 0$, comes from the interference between the two SFWM processes happening in the medium. We can directly observe the sidebands using the relation connecting the auto-correlation function of first and second order for a thermal light source~\cite{Loudon1983},
\begin{equation}
\overline{g}^{(2)}_{ii}(\tau) = 1 + |\overline{g}^{(1)}(\tau)|^2 \;\;\; (\text{for} \;\; i=1,2)\,,
\end{equation}
and calculating the spectrum from the Fast Fourier Transform $G_1 (\delta)$ of the first order correlation function $\overline{g}^{(1)}(\tau)$. Figure~\ref{fig:two_level_spectrum}(c) shows the corresponding $G_1 (\delta)$ for the auto-correlation curve of Fig.~\ref{fig:two_level_spectrum}(b), where we can see the large peak at $\delta = 0$ (corresponding to $\omega_s = \omega_l$) in the middle of symmetrical sidebands at $\omega_s = \pm \Delta$. 

\subsection{Filtering quantum correlations}

Our goal in filtering is to decrease the Rayleigh scattering enhancing the size of the oscillating structure on top of the plateau of Fig.~\ref{fig:two_level_spectrum}(b), i.e., to increase the proportion of the SFWM in the whole process. Of course, as part of the SFWM signal occurs with frequency $\omega_s = \omega_l$, the filtering will also alter the shape of the oscillating structure itself, decreasing the interference between its two components. The most characteristic trace of such decreased interference is the suppression of the minimum of $\overline{g}^{(2)}_{1,2}$ at $\tau = 0$. This suppression was most clearly demonstrated in the two-photon emission from single atoms in free space~\cite{Masters2023}, in which the amount of filtering of the $\omega_s = \omega_l$ component would control the statistics of light around $\tau = 0$, changing it from anti-bunching to bunching.

\begin{figure}[h!]
\centering
\includegraphics[width=7.5cm]{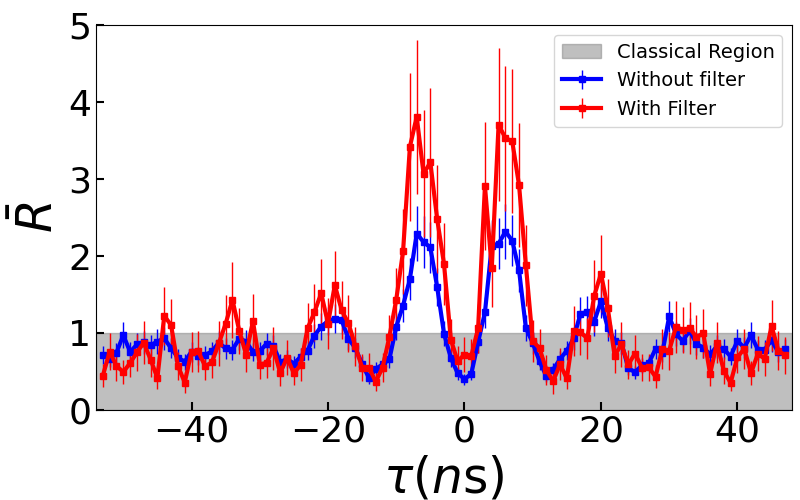}
\vspace{-0.3cm}
\caption{Cauchy--Schwarz parameter $\overline{R}(\tau)$ without (blue) and with (red) reflection-mode filtering. The data are shown in $1~\mathrm{ns}$ bins. The gray region is the classical domain $\overline{R}\leq1$. The data were acquired with $185~\mu\mathrm{W}$ in each pump beam and detuning $\Delta=8\Gamma$. The uncertainties were evaluated following the procedure used in Refs.~\onlinecite{PRL2022,OL2023}.}
\label{fig:Rbar_fbe}
\end{figure}

In this way, we aim to reject all components at $\omega_l$ and direct to our two detection regions only the components detuned from the excitation frequency, at $\omega_s \approx \omega_l \pm \Delta$, which present stronger quantum correlations~\cite{OL2023}. A typical result of our filtering procedure is shown in Fig.~\ref{fig:Rbar_fbe} for $\Delta = 8\Gamma$, where we plot $\overline{R}(\tau)$ with and without filtering. For filtering, each cavity is locked with its notch filter centered on $\omega_l$. The results without filtering were taken under the same conditions, just not locking the cavities. We can see then that the maximum of $\overline{R}(\tau)$ clearly increases with the filtering with respect to the curve without filtering. We can also see the minimum at $\tau = 0$ increasing, indicating a decreased interference at that point. The magnitudes of both effects are still small for what we need to move forward for the applications we have in mind, and below we are going to discuss the origin of such small amount of enhancement and how we plan to increase it.

Figure~\ref{fig:Rbar_fbe} uses a small $\delta \tau = 1\,$ns time window to acquire statistics, in order to better convey the system's time dynamics. The large error bars come specifically from the $\overline{g}^{(2)}_{22}(\tau = 0)$ value used in the calculation of all points in the $\overline{R}(\tau)$ curve, as the collection channel 2 presented a particularly large absorption and smaller statistics. In order to better quantify the filters' impact, then, we obtained all quantities to evaluate $\overline{R}(\tau_{max})$ with $\delta \tau = 3\,$ns, with $\tau_{max}$ the delay for the maximum value of $\overline{R}$. In this case, we obtain $\overline{R}(\tau_{max}) = 4.15 \pm 0.93$ with filters and $2.33 \pm 0.20$ without filters. The values of $\overline{g}^{(2)}_{ij}$ used for the calculation of $\overline{R}(\tau_{max})$ in each situation are given in Table~\ref{valuesRmax}.

\begin{table}[ht]
\centering
\caption{Correlation values used to evaluate $\overline{R}$ at the two peaks. We estimate each zero-delay autocorrelation by averaging the three $1~\mathrm{ns}$ bins centered at $\tau=-1,\,0,\,+1~\mathrm{ns}$. For the numerator of $\overline{R}$, we evaluate the cross-correlation products in Eqs.~(5) and~(6) in each bin and average them over two $3~\mathrm{ns}$ windows centered at $\tau=\pm6~\mathrm{ns}$. We use the same windows with and without filtering.}
\label{valuesRmax}
\begin{tabular}{@{}lll@{}}
\toprule
Quantity & with filters & without filters \\
\midrule
$\overline{g}^{(2)}_{11}(\tau = 0)$\hspace{0.8cm} & $1.87 \pm 0.15$ \hspace{0.8cm} & $1.85 \pm 0.06$ \\
$\overline{g}^{(2)}_{22}(\tau = 0)$ & $1.25 \pm 0.25$ & $1.53 \pm 0.11$ \\
$\langle\overline{g}^{(2)}_{12}\rangle_{\mathrm{avg}}$ & $3.11 \pm 0.11 $ & $2.57 \pm 0.04$ \\
\bottomrule
\end{tabular}

\vspace{0.2cm}
For $\langle\overline{g}^{(2)}_{12}\rangle_{\mathrm{avg}}$, the average is taken over the four detector pairs and the six bins in the two peak windows.
\end{table}

The filtering process implemented here follows a quite different principle from the one our group previously reported in Ref.~\onlinecite{OL2023}. This can be seen most directly on the symmetrical shape of the filtered curve in Fig.~\ref{fig:Rbar_fbe}. This is a result of both sidebands being equally directed to each detection channel, while the filtering in Ref.~\onlinecite{OL2023} directed a single sideband to each channel. Since the sidebands are emitted in a specific time order~\cite{Aspect1980}, the filtered curves in Ref.~\onlinecite{OL2023} were completely asymmetrical, with the quantum correlations appearing only for the direction of delays following the order of emission of the sidebands. The symmetry of the present process will be important for future demonstration of position-time entanglement between the two detection modes~\cite{Franson1989,Ou1990}. For the practical implementation of joint narrowband filters, this symmetry was also important for allowing the use of a single reference laser to lock both filters, making the long-term operation of the system more robust.

Another large difference with respect to Ref.~\onlinecite{OL2023} was the filter bandwidths. The large, 600-MHz bandwidth of the filters in Ref.~\onlinecite{OL2023} allowed for the passive stabilization of both filters, simultaneously, for the duration of an experimental run with decent statistics. However, such large bandwidth required large detunings to separate the components at $\omega_l \pm \Delta$ from the ones at $\omega_l$. The large detunings, on the other hand, resulted in smaller signal-to-noise ratios that ultimately led us to abandon that filtering strategy in favor of the one presented here. Once we change to smaller bandwidths, however, it becomes central to actively lock the filters at specific positions, which is a large portion of the challenge we addressed in the present work.

\subsection{Perspectives for improvement}

The level of filtering reported here is of the same order as the one reported in Ref.~\onlinecite{OL2023}. However, we understand that our new method is better suited for further improvement. We identified two main limitations now: (1) we still have a small signal-to-noise ratio; and (2) the filter bandwidth is too narrow, requiring optimization. The small signal, even as we work much closer to resonance, comes from a specific technical issue: the need to use intensity modulators right before the SPADs. The role of the intensity modulators is to prevent spurious light from the auxiliary reference laser from reaching the SPADs outside the measurement window, when the reference laser is on. The problem is the saturation of the SPADs even when not gated and its slow recovery, which leads to a much higher noise level during measurement. We see two possible solutions to be implemented in the future in order to remove the intensity modulators from the setup. One is to replace the linear detector for the light transmitted by the Fabry-P\'erot with another SPAD, and work with a reference laser at the single photon level to avoid saturating the SPADs used for the signals. Another option is to change the present SPADs for Superconducting Nanowire Single-Photon Detectors (SNSPDs), which are less sensitive and have a faster recovery from saturation by spurious light.

For our second limitation, our narrowband filters were designed to suppress scattered light of bandwidth on the order of half the natural linewidth of the transition or smaller, as expected for the central component of the measured spectra of the scattered light [as in Fig.~\ref{fig:two_level_spectrum}(c)] or from theoretical analysis of the Mollow triplet~\cite{Mollow1969,Masters2023}. We also expected to have the rate $R_R$ of linear single-photon scattering to be the main contribution for this central peak at $\omega_l$. However, even though our narrow reference field shows approximately $70\%$ rejection at the bottom of the notch filter, the single-photon detection rate decreases only by about $30\%$. This smaller reduction is consistent with the central component of the spectrum being broader than the cavity notch filter, so that only part of the collected light is rejected. We also note that changing slightly the position of the filter, around what should be the maximum transmission of the scattered light, does not change this value of 30\%, indicating a central component of the spectrum larger than we expected. We still do not fully understand the reason for this behavior. However, we plan to test this hypothesis by changing the filter cavity bandwidth and checking if the decrease in single-photon detections follows as expected. A possible way to do that in our system is by changing the cavity mirrors.

\section{Conclusion}\label{conclusion}

We have implemented a narrowband filtering architecture based on two actively stabilized Fabry--P\'erot cavities operating as reflection-mode notch filters, one in each photon-collection channel. The central feature of the instrument is the separation of the locking and filtering functions: an AOM-dithered auxiliary field detected in transmission provides the frequency discriminator, while the reflected output carries the detuned quantum light to the single-photon detectors. In contrast with conventional cavity-transmission filtering, the desired photons need not be resonant with a narrow cavity mode. Temporal multiplexing allows the reference and quantum fields to share the same optical path without exposing the detectors to the auxiliary field during photon counting or using the generated photons as a locking signal. The two independent filtering channels have comparable FSRs and linewidths and remain operational beyond the duration required for a typical correlation acquisition. When applied to the two-level-atom SFWM source, the filters preserve the observed violation of the Cauchy--Schwarz bound, with a higher central value of the correlation parameter than in the unfiltered measurement. Together with the narrow-field characterization, the accompanying reduction in the single-detector count rates shows that the notch filters remain effective under the conditions of the photon-pair experiment. These results establish actively stabilized reflection-mode notch filtering as a practical approach for rejecting narrow resonant spectral components in experiments with narrowband correlated light emitted in multiple directions.

\begin{acknowledgments}
This work was supported by the Conselho Nacional de Desenvolvimento Científico e Tecnológico (CNPq) through INCT-IQ (Grant No.~465469/2014-0) and CNPq/MCTI Call No.~26/2023 (Grant No.~422306/2023-1), by the Coordenação de Aperfeiçoamento de Pessoal de Nível Superior (CAPES/PROEX, Grant No.~23038.003069/2022-87), by the Fundação de Amparo à Ciência e Tecnologia do Estado de Pernambuco (FACEPE), and by the Fundação de Amparo à Pesquisa do Estado de São Paulo (FAPESP, Grant No.~2021/06535-0). The authors also acknowledge support from MCTI/FINEP/FNDCT through the Centros Temáticos 2023 program (QUANTA-UFPE, Project No.~1020/24) and from the U.S. Office of Naval Research (ONR, Grant No.~N62909-23-1-2014). A.~A.~C. de Almeida and D.~Felinto acknowledge fellowship support from Petrobras (Project No.~2025/00209-4).
\end{acknowledgments}

\section*{Author Declarations}

\subsection*{Conflict of Interest}
The authors have no conflicts to disclose.

\subsection*{Ethics Approval}
Ethics approval was not required for this study.

\subsection*{Author Contributions}
\textbf{G.~C. Borges} (Investigation, Formal analysis, Writing -- original draft).
\textbf{M. Duclerc} (Investigation, Formal analysis, Writing -- original draft).
\textbf{G. Palacios} (Formal analysis, Software).
\textbf{A.~A.~C. de Almeida} (Investigation, Writing -- original draft, Writing -- review \& editing).
\textbf{N.~R. de Melo} (Investigation, Writing -- review \& editing).
\textbf{D. Felinto} (Supervision, Formal analysis, Writing -- review \& editing).

\section*{Data Availability Statement}
The data that support the findings of this study are available from the corresponding author upon reasonable request.


\begin{thebibliography}{15}

\bibitem{Duan2001}
L.-M.~Duan, M.~D.~Lukin, J.~I.~Cirac, and P.~Zoller,
``Long-distance quantum communication with atomic ensembles and linear optics,''
\textit{Nature} \textbf{414}, 413--418 (2001).

\bibitem{Chou2007}
C.-W.~Chou, J.~ Laurat, H.~Deng, K.~S.~Choi, H.~de~Riedmatten, D.~Felinto, and H.~J.~Kimble,
``Functional Quantum Nodes for Entanglement Distribution over Scalable Quantum Networks,''
\textit{Science} \textbf{316}, 1316--1320 (2007).

\bibitem{Kimble2008}
H.~J.~Kimble,
``The quantum internet,''
\textit{Nature} \textbf{453}, 1023--1030 (2008).

\bibitem{Azuma2023}
K.~Azuma, S.~E.~Economou, D.~Elkouss, P.~Hilaire, L.~Jiang, H.-K. Lo, and I. Tzitrin,
``Quantum repeaters: From quantum networks to the quantum internet,''
\textit{Rev. Mod. Phys.} \textbf{95}, 045006 (2023).

\bibitem{Craddock2024}
A.~N.~Craddock, A.~Lazenby, G.~B.~Portmann, R.~Sekelsky, M.~Flament, and M.~Namazi,
``Automated Distribution of Polarization-Entangled Photons Using Deployed New York City Fibers,''
\textit{PRX Quantum} \textbf{5}, 030330 (2024).

\bibitem{Laurat2006}
J.~Laurat, H.~de~Riedmatten, D.~Felinto, C.-W.~Chou, E.~W.~Schomburg, and H.~J.~Kimble,
``Efficient retrieval of a single excitation stored in an atomic ensemble,''
\textit{Opt. Express} \textbf{14}, 6912--6918 (2006).

\bibitem{Mendes2011}
M.~S.~Mendes and D.~Felinto,
``Perspectives for laboratory implementation of the Duan-Lukin-Cirac-Zoller protocol for quantum repeaters,''
\textit{Phys. Rev. A} \textbf{84}, 062303 (2011).

\bibitem{PRL2022}
M.~O.~Ara\'ujo, L.~S.~Marinho, and D.~Felinto,
``Observation of nonclassical correlations in biphotons generated from an ensemble of pure two-level atoms,''
\textit{Phys. Rev. Lett.} \textbf{128}, 083601 (2022).

\bibitem{Marinho2025}
L.~S.~Marinho, M.~O.~Ara\'ujo, and D.~Felinto,
``Quantum and classical correlations in four-wave mixing from cold ensembles of two-level atoms,''
\textit{Phys. Rev. A} \textbf{111}, 013703 (2025).

\bibitem{deAlmeida2016}
A. J. F. de Almeida, M.-A. Maynard, C. Banerjee, D. Felinto, F. Goldfarb, and J. W. R. Tabosa,
``Nonvolatile optical memory via recoil-induced resonance in a pure two-level system,''
\textit{Phys. Rev. A} \textbf{94}, 063834 (2016).

\bibitem{Capella2022}
J. C. C. Capella, A. M. G. de Melo, J. P. Lopez, J. W. R. Tabosa, and D. Felinto,
``Atomic memory based on recoil-induced resonances,''
\textit{Phys. Rev. A} \textbf{106}, 013101 (2022).

\bibitem{deOliveira2015}
R. A. de Oliveira, G. C. Borba, W. S. Martins, S. Barreiro, D. Felinto, and J. W. R. Tabosa,
``Nonlinear optical memory for manipulation of orbital angular momentum of light,''
\textit{Opt. Lett.} \textbf{40}, 4939 (2015).

\bibitem{Borba2017}
G. C. Borba, D. Felinto, and J. W. R. Tabosa,
``Nonlinear optical memory: theory and experiment,''
\textit{J. Opt. Soc. Am. B} \textbf{34}, 2528 (2017).

\bibitem{Mollow1969}
B. R. Mollow,
``Power Spectrum of Light Scattered by Two-Level Systems,''
\textit{Phys. Rev.} \textbf{188}, 1969 (1969).

\bibitem{CohenTannoudji1979}
C. Cohen-Tannoudji and S. Reynaud,
``Atoms in strong light-fields: photon antibunching in single atom fluorescence,''
\textit{Phil. Trans. R. Soc. Lond. A} \textbf{293}, 223 (1979).

\bibitem{Aspect1980}
A. Aspect, G. Roger, S. Reynaud, J. Dalibard, and C. Cohen-Tannoudji,
``Time Correlations between the Two Sidebands of the Resonance Fluorescence Triplet,''
\textit{Phys. Rev. Lett.} \textbf{45}, 617 (1980).

\bibitem{Prasad2020}
A. S. Prasad, J. Hinney, S. Mahmoodian, K. Hammerer, S. Rind, P. Schneeweiss, A. S. S\text{\o}rensen, J. Volz, and A. Rauschenbeutel,
``Correlating photons using the collective nonlinear response of atoms weakly coupled to an optical mode,''
\textit{Nat. Photon.} \textbf{14}, 719 (2020).

\bibitem{Cordier2023}
M. Cordier, M. Schemmer, P. Schneeweiss, J. Volz, and A. Rauschenbeutel,
``Tailoring Photon Statistics with an Atom-Based Two-Photon Interferometer,''
\textit{Phys. Rev. Lett.} \textbf{131}, 183601 (2023).

\bibitem{Masters2023}
L. Masters, X.-X. Hu, M. Cordier, G. Maron, L. Pache, A. Rauschenbeutel, M. Schemmer, and J. Volz,
``On the simultaneous scattering of two photons by a single two-level atom,''
\textit{Nat. Photon.} \textbf{17}, 972 (2023).

\bibitem{Liu2024}
S. Liu, O. A. D. Sandberg, M. L. Chan, B. Schrinski, Y. Anyfantaki, R. B. Nielsen, R. G. Larsen, A. Skalkin, Y. Wang, L. Midolo, S. Scholz, A. D. Wieck, A. Ludwig, A. S. S\text{\o}rensen, A. Tiranov, P. Lodahl,
``Violation of Bell inequality by photon scattering on a two-level emitter,''
\textit{Nat. Phys.} \textbf{20}, 1429 (2024).

\bibitem{Maron2026}
G. Maron, A. B\text{\"o}lian, X.-X. Hu, L. Masters, A. Rauschenbeutel, and J. Volz,
``Few-MHz bandwidth tunable optical filter based on a fiber-ring resonator,''
\textit{Phys. Rev. Appl.} \textbf{26}, 024029 (2026).

\bibitem{Liu2024B}
Supplementary Material for Ref.~\onlinecite{Liu2024}.

\bibitem{OL2023}
L.~S.~Marinho, M.~O.~Ara\'ujo, W.~Martins, and D.~Felinto,
``Enhancing nonclassical correlations for light scattered by an ensemble of cold two-level atoms,''
\textit{Opt. Lett.} \textbf{48}, 3323--3326 (2023).

\bibitem{Drever1983}
R.~W.~P.~Drever, J.~L.~Hall, F.~V.~Kowalski, J.~Hough, G.~M.~Ford, A.~J.~Munley, and H.~Ward,
``Laser phase and frequency stabilization using an optical resonator,''
\textit{Appl. Phys. B} \textbf{31}, 97--105 (1983).

\bibitem{Black2001}
E.~D.~Black,
``An introduction to Pound--Drever--Hall laser frequency stabilization,''
\textit{Am. J. Phys.} \textbf{69}, 79--87 (2001).

\bibitem{Liu2020}
C.~Liu, Z.~Yue, Z.~Xu, M.~Ding, and Y.~Zhai,
``Far off-resonance laser frequency stabilization technology,''
\textit{Appl. Sci.} \textbf{10}, 3255 (2020).

\bibitem{Zugenmaier2018}
M. Zugenmaier, K. B. Dideriksen, A. S. S{\o}rensen, B. Albrecht, and E. S. Polzik,
``Long-lived non-classical correlations towards quantum communication at room temperature,''
\textit{Commun. Phys.} \textbf{1}, 76 (2018).

\bibitem{Zeng2021}
Y.~Zeng, Z.~Fu, Y.-Y.~Liu, X.-D.~He, M.~Liu, P.~Xu, X.-H.~Sun, and J.~Wang,
``Stabilizing a laser frequency by the Pound--Drever--Hall technique with an acousto-optic modulator,''
\textit{Appl. Opt.} \textbf{60}, 1159--1163 (2021).

\bibitem{Clauser1974}
J.~F.~Clauser,
``Experimental distinction between the quantum and classical field-theoretic predictions for the photoelectric effect,''
\textit{Phys. Rev. D} \textbf{9}, 853--860 (1974).

\bibitem{Loudon1983}
R. Loudon, {\it The Quantum Theory of Light} (Oxford University Press, New York, 1983).

\bibitem{Du2007PRL}
S.~Du, J.~Wen, M.~H.~Rubin, and G.~Y.~Yin,
``Four-wave mixing and biphoton generation in a two-level system,''
\textit{Phys. Rev. Lett.} \textbf{98}, 053601 (2007).

\bibitem{Wen2007PRA}
J.~Wen, S.~Du, and M.~H.~Rubin,
``Biphoton generation in a two-level atomic ensemble,''
\textit{Phys. Rev. A} \textbf{75}, 033809 (2007).

\bibitem{Franson1989}
J. D. Franson,
``Bell Inequality for Position and Time,''
\textit{Phys. Rev. Lett.} \textbf{62}, 2205 (1989).

\bibitem{Ou1990}
Z. Y. Ou, X. Y. Zou, L. J. Wang, and L. Mandel,
``Observation of Nonlocal Interference in Separated Photon Channels,''
\textit{Phys. Rev. Lett.} \textbf{65}, 321 (1990).

\end{thebibliography}
\end{document}